\documentclass{article}

\usepackage{arxiv}

\usepackage[utf8]{inputenc} % allow utf-8 input
\usepackage[T1]{fontenc}    % use 8-bit T1 fonts
\usepackage{hyperref}       % hyperlinks
\usepackage{url}            % simple URL typesetting
\usepackage{booktabs}       % professional-quality tables
\usepackage{amsfonts}       % blackboard math symbols
\usepackage{nicefrac}       % compact symbols for 1/2, etc.
\usepackage{microtype}      % microtypography
\usepackage{lipsum}		% Can be removed after putting your text content
\usepackage{graphicx}
\usepackage{doi}
\usepackage{amsmath}
\usepackage{authblk}
\usepackage[title]{appendix}
\usepackage[sort&compress,numbers]{natbib}
\title{Advantages of the adoption of a generalized flame displacement velocity as a central element of flamelet theory}

\author[ ~,1]{Hernan Olguin\thanks{Corresponding author\\ Email address: hernan.olguin@usm.cl}}%\corref{mycorrespondingauthor1}}
\author[2]{Pascale Domingo}
\author[2]{Luc Vervisch}
\author[3]{Christian Hasse}
\author[3]{Arne Scholtissek}

\affil[1]{Department of Mechanical Engineering, Universidad Técnica Federico Santa María,\linebreak Avenida España 1680, Valparaíso, Chile\linebreak}
\affil[2]{CORIA CNRS,  Normandie Universit\'e, INSA de Rouen, Technop\^ole du Madrillet, BP 8,\linebreak Saint-\'Etienne-du-Rouvray 76801, France\linebreak}
\affil[3]{Institute for Simulation of reactive Thermo-Fluid Systems, TU Darmstadt,\linebreak Otto-Berndt-Stra{\ss}e 2, 64287 Darmstadt, Germany}
\date{}

\begin{document}
\maketitle

\begin{abstract}
In combustion theory, flames are usually described in terms of the dynamics of iso-surfaces of a specific scalar. The flame displacement speed is then introduced as a local variable quantifying the progression of these iso-surfaces relative to the flow field. While formally defined as a scalar, the physical meaning of this quantity allows relating it with a vector pointing along the normal direction of the scalar iso-surface. In this work, this one-dimensional concept is extended by the introduction of a generalized flame displacement velocity vector, which is associated with the dynamics of iso-surfaces of two generic scalars, $\alpha$ and $\beta$. It is then shown how a new flamelet paradigm can be built around this velocity vector, which leads to a very compact and generic set of two-dimensional flamelet equations for thermochemical quantities and the conditioning scalar gradients, $g_{\alpha} = \lvert \nabla \alpha \rvert$ and $g_{\beta} = \lvert \nabla \beta \rvert$. The most important features of the developed framework are discussed in the context of partially-premixed flames, which provides significant insights into several aspects of the theory, including the nature of the different contributions to the flamelet equations for the conditioning scalar gradients and the fact that different flamelet coordinate systems (orthogonal and non-orthogonal) can be  characterized by the same flame displacement velocity vector. This approach opens an entire spectrum of possibilities for the definition of new two-dimensional composition spaces, which represents a very promising basis for the development of new variants of flamelet theory.
\end{abstract}

\vspace{15mm}

\section{Introduction}\label{introduction}

Flame-attached coordinate systems are an essential component of flamelet theory. Their formal introduction directly leads to well-defined transformation rules for the temporal and spatial derivatives appearing in the governing equations for the different (reactive) scalars of interest. This has been systematically used during the last decades to derive several sets of one-dimensional~\citep{Peters84,Pitsch98,Oijen00,Oijen01,Lodier11,vanOijen16,Xu13,Xuan14,Olguin13,Olguin14,Scholtissek15,Olguin15,Scholtissek17,Scholtissek18, Scholtissek19, Olguin19,Olguin23} and two-dimensional flamelet equations~\citep{Hasse05,Domingo08,Knudsen09,Nguyen10,Knudsen12,Mittal12,Mueller20,Scholtissek20,Novoselov21,Olguin232,Olguin24}.\\

Alternatively, flamelet theory admits a Lagrangian interpretation, which utilizes the velocity of the isosurfaces of the chosen conditioning scalars to characterize flamelet coordinate systems. For a one-dimensional flamelet description, this quantity coincides with Gibson's particle velocity~\citep{Gibson68} and it is associated with the dynamics of the isosurfaces of a single variable. For a two-dimensional flamelet transformation, on the other hand, two scalar isosurfaces need to be simultaneously considered, which leads to an extension of Gibson's velocity, the flamelet particle velocity, $\mathbf{u}_p$~\citep{Olguin232}. Given its definition, this velocity can be  mathematically related to the flame displacement speed, $s_d$, a widely-used variable in classical combustion theory~\citep{Williams85,Poinsot05}. Since the latter is typically defined in terms of the dynamics of the isosurfaces of the reaction progress variable, $Y_c$, the formalization of this connection is direct for a one-dimensional flamelet transformation adopting this scalar as main coordinate~\citep{Echekki99,Peters00}. For a two-dimensional flamelet formulation, on the other hand, the link between $\mathbf{u}_p$ and $s_d$ is not obvious when the reaction progress variable is not one of the chosen conditioning scalars. Despite the additional challenges, it has been recently shown in the context of an orthogonal two-dimensional coordinate system built in terms of the mixture fraction, $Z$, and a modified reaction progress variable, $\varphi$, how the flamelet particle velocity can be defined in such a way that its projection in the direction normal to the $Y_c$-isosurfaces recovers $s_d$~\citep{Olguin232}.

So far, the explicit use of the velocity of the flamelet coordinate system has allowed, among others, i) the development of different approaches based on the transport of flamelet structures in flow fields~\citep{Pitsch982,Pitsch00,Scholtissek17}, ii) the formulation of a coordinate-invariant set of one-dimensional flamelet equations for non-premixed flames~\citep{Klimenko01}, and iii) the derivation of a comprehensive and closed set of two-dimensional  flamelet equations for partially premixed flames in an orthogonal composition space~\citep{Olguin232}. These achievements clearly illustrate some of the advantages of the Lagrangian interpretation of the flamelet transformation. Unfortunately, the explicit use of $\mathbf{u}_p$ (or any equivalent quantity) in flamelet theory remains limited to the intermediate steps of the associated derivations only. As this work aims to demonstrate, this state of affairs is not optimal and it has consequences for the interpretability and applicability of the resulting theory. \\

The \textbf{main objective} of the present work is illustrating the advantages of reformulating flamelet theory in terms of the generalized flame displacement velocity, $\mathbf{s}_d$ (vector), which represents an extension of the classical flame displacement speed, $s_d$ (scalar). More specifically, it is shown that the adoption of $\mathbf{s}_d$ as a central element of the flamelet paradigm leads to:
\begin{itemize}
\item A simple procedure for the derivation of generalized flamelet equations, which is much more direct than the ones currently available in the literature.\\
\item A very compact and easy to interpret form of the flamelet equations for the conditioning scalar gradients, which comprises different effects in only few terms directly related with the projections of the flame displacement velocity.\\ 
\item The possibility of characterizing different composition space coordinate systems (orthogonal and non-orthogonal) through the same generalized flame displacement velocity, which allows the definition of new yet unexplored flamelet coordinate systems.\\
\end{itemize}

While the discussion is carried out in the context of partially premixed flames, the new framework utilizes a generic non-orthogonal two-dimensional coordinate system, ($\alpha$, $\beta$, $\tau$), and it is therefore suitable for use in other contexts too.

\section{The generalized flame displacement velocity}
We start summarizing different Lagrangian relations of relevance, which will allow us formally defining the generalized flame displacement velocity and obtaining a mathematical expression comprising all the conditions that this variable needs to satisfy in the context of a Lagrangian interpretation of flamelet theory. For this, we introduce a generic reactive scalar, $\psi$, two generic conditioning scalars (also called flamelet coordinates), $\alpha$ and $\beta$, and the corresponding scalar gradients, $g_{\alpha} = \lvert \nabla \alpha \rvert$ and $g_{\beta} = \lvert \nabla \beta \rvert$. In general, $\psi$ will represent a chemical species mass fraction or temperature, while $\alpha$ and $\beta$ will typically denote either i) a mixture fraction and some sort of reaction progress variable~\citep{Domingo08,Knudsen09,Nguyen10,Knudsen12,Scholtissek20,Olguin232}, ii) two different mixture fractions~\citep{Hasse05}, or iii) a mixture fraction and entalphy~\citep{Mittal12}. For all these variables, appropriate governing equations in physical space, ($\mathbf{x},t$), can be obtained by means of the following operator
\begin{equation}
    \frac{\partial (\cdot)}{\partial t} = - \mathbf{u}\cdot\nabla(\cdot) + \Gamma_{(\cdot)},
    \label{operator1}
\end{equation}
where $\mathbf{u}$ denotes the flow velocity and, depending on the exact nature of the variable under consideration, $\Gamma_{(\cdot)}$ will contain terms associated with different effects, such as diffusion, chemical reactions, change of phases, etc.  

Making use of the chain rule, the following operator can be defined~(\cite{Klimenko01,Scholtissek20,Olguin232})
\begin{equation}
\frac{\partial (\cdot)}{\partial \tau} = \frac{\partial (\cdot)}{\partial t} + \mathbf{u}_p \cdot \nabla (\cdot),
\label{operator2}
\end{equation}
where $\tau$ denotes a time-like variable and $\mathbf{u}_p$ represents the velocity of a mass-less (flamelet) particle traveling with the intersection between isosurfaces of $\alpha$ and $\beta$. This directly implies that $\partial (\cdot)/\partial \tau$ is calculated keeping $\alpha$ and $\beta$ constant. 

Combining now Eqs.~(\ref{operator1}) and (\ref{operator2}), we obtain
\begin{equation}
\mathbf{s}_d \cdot \mathbf{n}_{(\cdot)} = (\mathbf{u}_p - \mathbf{u})  \cdot \mathbf{n}_{(\cdot)} = \frac{1}{\lvert \nabla (\cdot) \rvert} \left(\frac{\partial (\cdot) }{\partial \tau} - \Gamma_{(\cdot)}\right), 
\label{sd-condition}
\end{equation}
where $\mathbf{s}_d = (\mathbf{u}_p - \mathbf{u})$ is defined as the generalized flame displacement velocity and $\mathbf{n}_{(\cdot)} = \nabla (\cdot)  / \lvert \nabla (\cdot) \rvert$ is a generic unit vector normal to the corresponding isosurface of the scalar to which it is applied. This equation provides an entire set of conditions that the generalized flame velocity needs to satisfy in the context of a Lagrangian interpretation of the flamelet formalism. In the next section, it will be shown how the  successive application of Eq.~(\ref{sd-condition}) allows obtaining a general set of flamelet equations. Before, it is remarked that, since Eq.~(\ref{operator2}) directly implies that $\partial \alpha/\partial \tau = \partial \beta / \partial \tau = 0$, the specialization of Eq.~(\ref{sd-condition}) for the flamelet coordinates, $\alpha$ and $\beta$, leads to the following specific conditions~(\cite{Olguin232})
\begin{equation}
   \mathbf{s}_d \cdot \mathbf{n}_{\alpha}= - \frac{\Gamma_{\alpha}}{\lvert \nabla \alpha \rvert}~\text{      }~\text{   and     }~\text{       }~\mathbf{s}_d \cdot \mathbf{n}_{\beta} = - \frac{\Gamma_{\beta}}{\lvert \nabla \beta \rvert}.
   \label{condition-alpha-beta}
\end{equation}

Finally, as pointed out in the introduction, we remark that the flame displacement velocity, $\mathbf{s}_d$, is indeed a generalization of the classical flame displacement speed, $s_d$. This relation can be easily seen by considering the particular case in which either $\alpha$ or $\beta$ corresponds to the reaction progress variable, $Y_c$. In such a situation, Eq.~(\ref{sd-condition}) directly yields $(\mathbf{s}_d \cdot \mathbf{n}_{c})  = - \Gamma_{c} / \lvert \nabla Y_c \rvert = - s_d$. It must be noted, however, that the possibility of establishing a relation between $\mathbf{s}_d$ and $s_d$ does not require $Y_c$ to be one of the flamelet coordinates, which will be explored in more detail later.

\section{Direct procedure for the derivation of general flamelet equations}
%\addvspace{10pt}
It will be shown now how the successive application of Eq.~(\ref{sd-condition}) allows developing a very direct procedure for the derivation of general flamelet equations for any variable governed by an equation of the form of Eq.~(\ref{operator1}). For this, we start re-arranging this equation in the following more convenient form
\begin{equation}
  \frac{\partial (\cdot)}{\partial \tau} = \Gamma_{(\cdot)} + \mathbf{s}_d \cdot \nabla (\cdot) .
  \label{operator-psi}
\end{equation}
Making use of the classical spatial flamelet transformation~\citep{Nguyen10} 
\begin{equation}
\nabla (\cdot) = \frac{\partial (\cdot)}{ \partial \alpha} \nabla \alpha +  \frac{\partial (\cdot)}{\partial \beta} \nabla \beta,
\label{transf}
\end{equation}
we can rewrite the second term at the right hand side (RHS) of Eq.~(\ref{operator-psi}) as
\begin{equation}
  \mathbf{s}_d \cdot \nabla (\cdot) =   \mathbf{s}_d \cdot \left( \lvert \nabla \alpha \rvert \frac{\partial (\cdot)}{\partial \alpha} \mathbf{n}_{\alpha} + \lvert \nabla \beta \rvert   \frac{\partial (\cdot)}{\partial \beta}  \mathbf{n}_{\beta} \right),
  \label{LHS-sd}
\end{equation}
which leads to
\begin{equation}
  \frac{\partial (\cdot)}{\partial \tau} = \Gamma_{(\cdot)} + (\mathbf{s}_d \cdot \mathbf{n}_{\alpha})  g_{\alpha} \frac{\partial (\cdot)}{\partial \alpha}   + (\mathbf{s}_d \cdot \mathbf{n}_{\beta})  g_{\beta} \frac{\partial (\cdot)}{\partial \beta}.
  \label{psi-sd-equation}
\end{equation}
In Eq.~(\ref{psi-sd-equation}), the projections of the generalized flame velocity in $\alpha$ and $\beta$ directions appear in an explicit way, which allows the direct introduction of Eq.~(\ref{condition-alpha-beta}) to obtain
\begin{equation}
  \frac{\partial (\cdot)}{\partial \tau} = \Gamma_{(\cdot)} - \Gamma_{\alpha} \frac{\partial (\cdot)}{\partial \alpha}   - \Gamma_{\beta} \frac{\partial (\cdot)}{\partial \beta}.
  \label{flamelet}
\end{equation}
This operator represents the aimed general flamelet equation governing the evolution of any variable that could be described by an equation of the form of Eq.~(\ref{operator1}). Interestingly, and even when the current derivation utilizes a generic non-orthogonal coordinate system, Eq.~(\ref{flamelet}) has exactly the same form of the  general flamelet equation derived by~\cite{Olguin232} for the case of an orthogonal coordinate system (see Eq.~(14) in that work). Despite this similarity, it must be noted that the specification and transformation of $\Gamma_{(\cdot)}$ will lead to different terms for non-orthogonal and orthogonal coordinate systems, as it is shown in Appendix~\ref{appendix1}. More importantly, it is remarked that the current procedure significantly differs from the one employed by~\cite{Olguin232}, since the latter requires previously determining a specific expression for $\mathbf{s}_d$ (see Eq.~(\ref{sd-1})), which needs then to be inserted in Eq.~(\ref{operator-psi}). This would involve an extensive amount of algebra for the non-orthogonal coordinate system adopted in this work. In contrast, the current formalism makes effective use of the relations comprised in Eq.~(\ref{sd-condition}). This results in a very simple and direct procedure, which highlights the advantages of the proposed Lagrangian paradigm.

\section{Compact flamelet equations for the gradients of the conditioning scalars}
%\addvspace{10pt}

The aim of this section is showing how the scalar gradients, $g_{\alpha}$ and $g_{\beta}$, can be explicitly connected to the different projections of the generalized flame displacement velocity in a very compact and easy to interpret form. During the last decades, several different scalar gradient equations have been derived in the literature in both physical~\citep{Mantel94,Kollmann98,Chakraborty05,Swaminathan05,Chakraborty08,Olguin15,Dopazo16,Sandeep18,Cifuentes18,Dopazo18,Yu19,Yu192} and composition space~\citep{Hasse05,Peters09,Scholtissek17,Scholtissek19,Olguin19,Olguin23}. Since the current analysis focus on the latter, we start considering the following balance equation for the gradient of a general scalar, $S$, which is governed by Eq.~(\ref{operator1})~\citep{Olguin232}
\begin{equation}
    \frac{\partial g_S} {\partial t} + \mathbf{u} \cdot \nabla g_S = g_S a_S + \frac{\partial \Gamma_S}{\partial n_S},
    \label{gs}
    \end{equation}
where $a_S = - \mathbf{n}_{S} \cdot \nabla \mathbf{u} \cdot \mathbf{n}_{S}$ represents the strain rate and $\partial (\cdot) / \partial n_S = \mathbf{n}_S \cdot \nabla (\cdot)$ denotes a directional derivative. After specializing Eq.~(\ref{operator1}) for $g_S$, a direct comparison with Eq.~(\ref{gs}) leads to $\Gamma_{g_S} = g_S a_S + \partial \Gamma_S / \partial n_S$, which can be used together with Eq.~(\ref{psi-sd-equation}) to obtain
\begin{equation}
\frac{\partial g_{S}}{\partial \tau} = a_{S} g_{S} + \frac{\partial \Gamma_{S}}{\partial n_{S}} + s_d^{\alpha}  g_{\alpha} \frac{\partial g_{S}}{\partial \alpha}   + s_d^{\beta}  g_{\beta} \frac{\partial g_{S}}{\partial \beta},
\label{g-physical2}
\end{equation}
where the flame velocity projections in $\alpha$ and $\beta$ directions, $s_d^{\alpha} = (\mathbf{s}_d \cdot \mathbf{n}_{\alpha})$ and $s_d^{\beta} = (\mathbf{s}_d \cdot \mathbf{n}_{\beta})$, have been introduced for compactness. While Eq.~(\ref{g-physical2}) is valid for both $g_{\alpha}$ and $g_{\beta}$, only the former will be considered in the rest of this work for clarity. The obtained results, however, are easily transferred to $g_{\beta}$.

Using now the definition of the directional derivative, the flamelet transformation (Eq.~(\ref{transf})) and Eq.~(\ref{condition-alpha-beta}), the second term at the RHS of Eq.~(\ref{g-physical2}) can be rewritten as
\begin{align}
\frac{\partial \Gamma_{\alpha}}{\partial n_{\alpha}}  & = \mathbf{n}_{\alpha} \cdot \left( \frac{\partial \Gamma_{\alpha}}{\partial \alpha} \nabla \alpha + \frac{\partial \Gamma_{\alpha} }{\partial \beta} \nabla \beta \right) \nonumber \\ & = - g_{\alpha} \frac{\partial \left(s_d^{\alpha} g_{\alpha}\right)}{\partial \alpha}  - g_{\beta} \frac{\partial \left(s_d^{\alpha} g_{\alpha}\right)}{\partial \beta} \Theta,
\end{align}
where $\Theta = (\mathbf{n}_{\alpha}\cdot \mathbf{n}_{\beta})$ represents a quantification of the degree of alignment between the unit vectors associated with the corresponding iso-surfaces of $\alpha$ and $\beta$.
With this, Eq.~(\ref{g-physical2}) becomes
\begin{equation}
\frac{\partial g_{\alpha}}{\partial \tau} 
= \underbrace{a_{\alpha} g_{\alpha}\phantom{\bigg|}}_{\mathrm{I}} 
\underbrace{- g_{\alpha}^2 \frac{\partial s_d^{\alpha}}{\partial \alpha}\phantom{\bigg|}}_{\mathrm{II}} 
+ \underbrace{s_d^{\beta} g_{\beta} \frac{\partial g_{\alpha} }{\partial \beta}\phantom{\bigg|}}_{\mathrm{III}}  
\underbrace{- g_{\beta} \frac{\partial (s_d^{\alpha} g_{\alpha})}{\partial \beta} \Theta\phantom{\bigg|}}_{\mathrm{IV}},
\label{g-sd}
\end{equation}
which represents the aimed compact formulation of the $g_{\alpha}$-flamelet equation explicitly connecting this quantity with the projections of the generalized flame velocity, $s_d^{\alpha}$ and $s_d^{\beta}$. The four terms at the RHS of Eq.~(\ref{g-sd}) can be classified into two different kinds of contributions. First, terms I and II correspond to effects normal to the $\alpha$-isosurface, where term I is directly attributable to strain and term II will include the different contributions contained in $\Gamma_{\alpha}$. In general, these comprise diffusion, curvature and chemical reactions. Terms III and IV, on the other hand, correspond to tangential effects, where the former can be formally interpreted as convection in composition space and the latter comprises the effects associated with the non-orthogonality of the ($\alpha$,$\beta$) space.

Of course, Eq.~(\ref{g-sd}) is not the only form that the $g_{\alpha}$-equation can take, and, depending on the context, alternative formulations can be more convenient. For example, term IV can be rewritten as
\begin{equation}
   - g_{\beta} \frac{\partial (s_d^{\alpha} g_{\alpha})}{\partial \beta} \Theta =  - g_{\beta}  s_d^{\alpha} \Theta \frac{\partial g_{\alpha}}{\partial \beta} - g_{\alpha} g_{\beta} \Theta \frac{\partial s_d^{\alpha}}{\partial \beta}, 
   \label{splitIV}
\end{equation}
where the first term at the RHS corresponds to a further contribution to convection, while the second one is clearly depending on the cross scalar dissipation rate, $\chi_{\alpha,\beta} = g_{\alpha} g_{\beta} \Theta$. Making use of Eq.~(\ref{splitIV}), Eq.~(\ref{g-sd}) yields
\begin{equation}
\frac{\partial g_{\alpha}}{\partial \tau} = a_{\alpha} g_{\alpha} - g_{\alpha}^2 \frac{\partial s_d^{\alpha}}{\partial \alpha} + (s_d^{\beta} - s_d^{\alpha} \Theta) g_{\beta} \frac{\partial g_{\alpha} }{\partial \beta} - g_{\beta} g_{\alpha} \frac{\partial s_d^{\alpha}}{\partial \beta} \Theta.
\label{g-sd-2}
\end{equation}
Also, this equation could be further modified by replacing the strain rate, $a_{\alpha}$, by the stretch rate
\begin{equation}
K_{s,\alpha} = \frac{1}{\rho} \frac{\partial (\rho s_d^{\alpha})}{\partial n_{\alpha}},
\end{equation}
as it is typically done for premixed flames~(see for example~\cite{DeGoey97,DeGoey99,Scholtissek18,Scholtissek19,Olguin23}). It has been recently shown that this  eliminates the temporal derivatives appearing in the $g$-equation, but the resulting formulation would still be able to capture unsteady effects~\citep{Olguin23}. Probably, the most important advantage of the consideration of $K_s$ is its suitability for the study of unstretched flames,  for which $K_{s,\alpha} = 0$. 

To obtain a better understanding of the derived equations, we consider now a two-dimensional orthogonal coordinate system, which implies $\Theta = 0$ and reduces Eqs.~(\ref{g-sd}) and (\ref{g-sd-2}) to
\begin{equation}
\frac{\partial g_{\alpha}}{\partial \tau} = s_d^{\beta} g_{\beta} \frac{\partial g_{\alpha} }{\partial \beta} + a_{\alpha} g_{\alpha}  - g_{\alpha}^2 \frac{\partial s_d^{\alpha}}{\partial \alpha}.
\label{g-sd-orthogonal}
\end{equation}
Here, the only term including effects in the $\beta$-coordinate is the convective term, $s_d^{\beta} g_{\beta} \partial g_{\alpha} / \partial \beta$, which must therefore comprise all deviations from a classical one-dimensional flamelet formulation. The fact that these deviations are proportional to $s_d^{\beta}$ implies that a one-dimensional description is asymptotically recovered when the flame displacement velocity tends to align with $\nabla \alpha$, since $(\mathbf{s}_d \cdot \mathbf{n}_{\beta}) \to 0$ in that limit. On the contrary, when the flame displacement velocity tends to align with $\nabla \beta$,  the convective term is expected to become dominant. 

This analysis is consistent with results recently reported in the context of the orthogonal coordinate system, $(Z,\varphi)$~\citep{Olguin232}. In that work, a set of initially non-interacting premixed flamelet structures was perturbed by the introduction of a moderate amount of stratification in $Z$-direction (small values of $a_Z$), establishing a two-dimensional partially premixed steady flamelet in which premixed-like effects were expected to be dominant. The budgets of the $g_{\varphi}$-equation showed a very limited contribution of the tangential convection, indicating a quasi one-dimensional situation. For the $g_Z$-equation, on the other hand, it was found that diffusion in $Z$-direction was mainly balanced by convection in $\varphi$-direction, a clear consequence of the dominance of the premixed effects taking place in the studied case. 

Finally, it is noted that bringing Eq.~(\ref{g-sd-orthogonal}) into the same form analyzed by~\cite{Olguin232} requires the expansion of $s_d^{\alpha}$ and $s_d^{\beta}$ into their three major contributions: Diffusion, curvature and a source term. For example, if the $\Gamma$-terms appearing in Eq.~(\ref{operator1}) are specified as
\begin{equation}
   \Gamma_{(\cdot)} = \frac{1}{\rho} \nabla \cdot \left(\rho D \nabla (\cdot) \right) + \frac{\dot{S}_{(\cdot)}}{\rho},
   \label{gamma}
\end{equation}
where $\rho$ is the density, $D$ is a diffusion coefficient, and $\dot{S}_{(\cdot)}$ is a generic source term (which can contain chemical reactions, electric fields, other sources associated with change of phases, etc), Eq.~(\ref{condition-alpha-beta}) allows expressing $s_d^{\alpha}$ and $s_d^{\beta}$ as  
\begin{equation}
    s_d^{\alpha} = - \frac{1}{\rho} \frac{\partial }{\partial \alpha} \left( \rho D g_{\alpha} \right) + D \kappa_{\alpha} - \frac{\dot{S}_{\alpha}}{\rho g_{\alpha}}~\text{      }~\text{   and     }~\text{       }~s_d^{\beta} = - \frac{1}{\rho} \frac{\partial }{\partial \beta} \left( \rho D g_{\alpha} \right) + D \kappa_{\beta} - \frac{\dot{S}_{\beta}}{\rho g_{\beta}},
    \label{sd-99}
\end{equation}
respectively. Here, $\kappa_{\alpha} = - \nabla \cdot \mathbf{n}_{\alpha}$ and $\kappa_{\beta} = - \nabla \cdot \mathbf{n}_{\beta}$ are the curvatures of the $\alpha$ and $\beta$-isosurfaces and $\mathbf{n}_{\alpha} = \nabla \alpha / \lvert \nabla \alpha \rvert$ and $\mathbf{n}_{\beta} = \nabla \beta / \lvert \nabla \beta \rvert$ represent two unit vectors. Evidently, inserting Eq.~(\ref{sd-99}) into Eq.~(\ref{g-sd-orthogonal}) would lead to several new terms and a considerably less compact formulation of the $g$-equation, which highlights the advantages of the current approach.

\section{The generalized flame displacement velocity as an unifying element between coordinate systems}
\label{Family}
%\addvspace{10pt}

We proceed to show now how $\mathbf{s}_d$ can be used to connect the generic non-orthogonal reference frame, ($\alpha, \beta$), to a new \textit{orthogonal} coordinate system, ($\alpha, \beta'$), where $\beta'$ represents a modification of $\beta$ satisfying $\nabla \alpha \cdot \nabla \beta' = 0$. The formalism requires that $\alpha$ and $\beta$ are well-defined quantities, in the sense that $\Gamma_{\alpha}$ and $\Gamma_{\beta}$ must be known functions (which is satisfied by the mixture fraction and most reaction progress variables proposed in the literature). Based on this, the approach allows obtaining an appropriate definition for $\Gamma_{\beta'}$ ensuring that both coordinate systems are characterized by the same generalized flame displacement velocity. 

We start the derivation obtaining an expression for $\mathbf{s}_d$ characterizing the ($\alpha,\beta$) coordinate system. For this, we decompose this quantity as a combination of the unit vectors $\mathbf{n}_{\alpha}$ and $\mathbf{n}_{\beta}$. This yields
\begin{align}
  \mathbf{s}_{d} = f_1   \mathbf{n}_{\alpha} + f_2 \mathbf{n}_{\beta},
  \label{deco1}
\end{align}
where $f_1$, $f_2$ are functions to be determined. Multiplying Eq.~(\ref{deco1}) by $\mathbf{n}_{\alpha}$ and $\mathbf{n}_{\beta}$ (dot products), and making use of Eq.~(\ref{condition-alpha-beta}), we obtain
\begin{equation}
f_1 + f_2 \Theta = -\frac{\Gamma_{\alpha}}{\lvert \nabla \alpha \rvert}~\text{      }~\text{   and     }~\text{       }~f_1 \Theta + f_2 = -\frac{\Gamma_{\beta}}{\lvert \nabla \beta \rvert},
\label{f1-f2}
\end{equation}
respectively. Further, since Eq.~(\ref{f1-f2}) represents a closed system with two equations and two unknowns, $f_1$ and $f_2$ can be directly determined and replaced into Eq.~(\ref{deco1}) to obtain
\begin{equation}
    \mathbf{s}_d = \frac{1}{(1 - \Theta^2)} \left(\left[ \frac{\Gamma_{\beta}}{\lvert \nabla \beta \rvert} \Theta - \frac{\Gamma_{\alpha}}{\lvert \nabla \alpha \rvert} \right] \mathbf{n}_{\alpha}+ \left[ \frac{\Gamma_{\alpha}}{\lvert \nabla \alpha \rvert} \Theta - \frac{\Gamma_{\beta}}{\lvert \nabla \beta \rvert} \right] \mathbf{n}_{\beta} \right).
    \label{sd-1}
\end{equation}
Similarly, the above-presented procedure can be used to obtain an expression for the flame velocity characterizing the ($\alpha,\beta'$) coordinate system. More specifically, since for this particular situation $\Theta = 0$, this yields
\begin{equation}
\mathbf{s}_d = - \frac{\Gamma_{\alpha}}{\lvert \nabla \alpha \rvert} \mathbf{n}_{\alpha} - \frac{\Gamma_{\beta'}}{\lvert \nabla \beta' \rvert} \mathbf{n}_{\beta'},
\label{sd-2}
\end{equation}
which is consistent with the corresponding expression obtained by~\cite{Olguin232} for an orthogonal coordinate system in the context of partially premixed flames. 

Now, in order to introduce a formal connection between $\beta$ and $\beta'$, a further condition needs to be imposed, namely that Eqs.~(\ref{sd-1}) and (\ref{sd-2}) represent the same physical entity (even when written in terms of combinations of different vectors). This directly implies
\begin{equation}
\mathbf{s}_d \cdot \mathbf{n}_{\beta} = -\frac{\Gamma_{\alpha}}{\lvert \nabla \alpha \rvert} (\mathbf{n}_{\alpha} \cdot \mathbf{n}_{\beta})  -\frac{\Gamma_{\beta'}}{\lvert \nabla \beta' \rvert} (\mathbf{n}_{\beta'} \cdot \mathbf{n}_{\beta}) = -\frac{\Gamma_{\beta}}{\lvert \nabla \beta \rvert},
\label{projection}
\end{equation}
where use of Eqs.~(\ref{sd-2})~and~(\ref{condition-alpha-beta}) has been made. This equation allows to obtain an explicit relation for $\Gamma_{\beta'}$ ensuring that the condition $\nabla \alpha \cdot \nabla \beta' = 0$ is satisfied. For this, the products ($\mathbf{n}_{\alpha} \cdot \mathbf{n}_{\beta}$) and ($\mathbf{n}_{\beta'} \cdot \mathbf{n}_{\beta}$) must be first reduced, which can be done making use of the definition of $\mathbf{n}_{\beta}$ and the fact that the directional derivatives can be expressed in terms of the ($\alpha,\beta'$) coordinate system as $\partial \beta / \partial n_{(\cdot)} = \lvert \nabla (\cdot) \rvert \partial \beta / \partial (\cdot)$. This leads to
\begin{equation}
    (\mathbf{n}_{\alpha} \cdot \mathbf{n}_{\beta}) = \frac{\lvert \nabla \alpha \rvert}{\lvert \nabla \beta \rvert}\frac{\partial \beta}{\partial \alpha}~\text{      }~\text{   and     }~\text{       }~(\mathbf{n}_{\beta'} \cdot \mathbf{n}_{\beta}) = \frac{\lvert \nabla \beta' \rvert}{\lvert \nabla \beta \rvert}\frac{\partial \beta}{\partial \beta'},
\end{equation}
which can be replaced in Eq.~(\ref{projection}) to obtain
\begin{equation}
    \Gamma_{\beta'} = \frac{\Gamma_{\beta} - \Gamma_{\alpha} \frac{\partial \beta}{\partial \alpha}}{\frac{\partial \beta}{\partial \beta'}}.
\end{equation}
This allows writing the aimed $\beta'$ balance equation as
\begin{equation}
    \frac{\partial \beta'}{\partial t} + \mathbf{u} \cdot \nabla \beta' = \frac{\Gamma_{\beta} - \Gamma_{\alpha} \frac{\partial \beta}{\partial \alpha}}{\frac{\partial \beta}{\partial \beta'}}.
    \label{beta'-2}
\end{equation}
It should be noted at this point that Eq.~(\ref{beta'-2}) is indeed a generalization of the $\varphi$-equation introduced by~\cite{Olguin232}. Despite the similarities, the procedure presented in this work importantly differs from the one used in~\citep{Olguin232}.\\

In order to clarify possible applications of the procedure described in this section, we will focus the remainder of this analysis on the particular case of partially premixed combustion (without loss of generality), which is naturally described by the mixture fraction, $Z$, and the reaction progress variable, $Y_c$, a coordinate system that has been studied by~\cite{Nguyen10}. As pointed out before, a modified reaction progress variable, $\varphi$, has been recently proposed, which satisfies the condition $\nabla Z \cdot \nabla \varphi = 0$ (orthogonality)~\citep{Olguin232}. This variable allows defining the ($Z, \varphi$)-space, which, according to the analysis above, is connected with the original ($Z, Y_c$)-space through the generalized flame displacement velocity. We will consider now the third member of the ($Z, Y_c$) family of coordinates, namely the orthogonal space built by $Y_c$ and a modified mixture fraction, which will be denoted as $\zeta$. This coordinate system is a very interesting one, since it allows recovering the asymptotic limits of non-premixed and premixed combustion in a very novel way, which will be explained in detail below. 

We start the analysis noting that, in terms of ($Y_c, \zeta$), the flamelet transformation becomes
\begin{equation}
    \nabla (\cdot) = \frac{\partial (\cdot)}{\partial Y_c} \nabla Y_c + \frac{\partial (\cdot)}{\partial \zeta} \nabla \zeta.
    \label{flameletycxi}
\end{equation}
Now, in order to formally establish the connection between $\zeta$ and $Z$, the condition $\partial Z / \partial \zeta = 1$ is introduced. This is analogous to the condition imposed by~\cite{Olguin232} for the reaction progress variable, $\partial Y_c / \partial \varphi = 1$, and directly implies
\begin{equation}
    \nabla \zeta = \nabla Z - (\mathbf{n}_c \cdot \nabla Z) \mathbf{n}_c.
    \label{nablaxi}
\end{equation}
In other words, Eq.~(\ref{nablaxi}) shows that $\nabla \zeta$ corresponds to the component of $\nabla Z$ orthogonal to $\nabla Y_c$. For premixed combustion in the context of unity Lewis number, $\nabla Z \to 0$, which implies that $\nabla \zeta \to 0$ in this asymptotic limit. Therefore, as expected,  the flamelet transformation reduces to $\nabla(\cdot)\approx \partial (\cdot) / \partial Y_c \nabla Y_c$. In the non-premixed limit, on the other hand, the gradients of the mixture fraction and the reaction progress variable tend to align, which means $(\mathbf{n}_Z \cdot \mathbf{n}_c) \to 1$. This implies that $\nabla \zeta \to 0$, which, surprisingly, leads to the same reduced transformation rule for $\nabla (\cdot)$ obtained for premixed combustion. This very interesting feature represents a major motivation for the further study of this new coordinate system. Unfortunately, the adoption of the ($Y_c, \zeta$)-space also introduces some issues, which  render its full exploration difficult. These, however, are solved by the approach introduced in this section, as it will be discussed next. 

The first challenge in the implementation of the ($Y_c, \zeta$) coordinate system, is the lack of a specific expression for $\Gamma_{\zeta}$, which can be now obtained by setting $\alpha = Y_c$ and $\beta = Z$ in Eq.~(\ref{beta'-2}). This yields
\begin{align}
    \Gamma_{\zeta} & = \Gamma_{Z} - \Gamma_{c} \frac{\partial Z}{\partial Y_c},
    \label{gammaxi}
\end{align}
where the condition $\partial Z / \partial \zeta = 1$ has been introduced and where $\Gamma_Z$ and $\Gamma_c$ can be specified by means of Eq.~(\ref{gamma}) with $\dot{S}_Z = 0$ and $\dot{S}_c = \dot{\omega}_c$, respectively. 

Making use of the flamelet transformation, Eq.~(\ref{flameletycxi}), $\Gamma_Z$ can be further worked out as
\begin{align}
\Gamma_{Z} & = \frac{1}{\rho} \nabla \cdot \left( \rho D \left( \frac{\partial Z}{\partial Y_c} \nabla Y_c + \nabla \zeta \right) \right) \\ &= \frac{1}{\rho} \left[ \rho D g_c^2 \frac{\partial^2 Z}{\partial Y_c^2} + \frac{\partial Z}{\partial Y_c} \nabla \cdot (\rho D \nabla Y_c ) + \nabla \cdot (\rho D \nabla \zeta )  \right], \nonumber 
\end{align}
which allows rewriting Eq.~(\ref{gammaxi}) as
\begin{equation}
  \Gamma_{\zeta} = \frac{1}{\rho} \nabla \cdot (\rho D \nabla \zeta) + \frac{\dot{\omega}_{\zeta}}{\rho},
\end{equation}
where
\begin{equation}
\dot{\omega}_{\zeta} = \rho D g_c^2 \frac{\partial^2 Z}{\partial Y_c^2} - \frac{\partial Z}{\partial Y_c} \dot{\omega}_c.
\end{equation}
As shown in Appendix~\ref{appendix1}, the $\Gamma$-terms in Eq.~(\ref{flamelet}) can be explicitly written in terms of the coordinates $\alpha$ and $\beta$, which can be then specialized for $Y_c$ and $\zeta$ to obtain
\begin{align}
    \rho \frac{\partial \psi}{\partial \tau} & = \rho D g_c^2 \frac{\partial^2 \psi}{\partial Y_c^2} + \rho D g_{\zeta}^2 \frac{\partial^2 \psi}{\partial \zeta^2} + \dot{S}_{\psi} - \dot{\omega}_c \frac{\partial \psi}{\partial Y_c} \nonumber \\ & - \left( \rho D g_c^2 \frac{\partial^2 Z}{\partial Y_c^2} -  \dot{\omega}_c \frac{\partial Z}{\partial Y_c}\right) \frac{\partial \psi}{\partial \zeta},
\end{align}
where the corresponding scalar gradients are given by 
\begin{equation}
    \frac{\partial g_c}{\partial \tau} = s_d^{\zeta} g_{\zeta} \frac{\partial g_c}{\partial \zeta} + a_{c} g_c - g_c^2 \frac{\partial s_d^c}{\partial Y_c}
\end{equation}
and
\begin{equation}
    \frac{\partial g_{\zeta}}{\partial \tau} = s_d^{c} g_{c} \frac{\partial g_{\zeta}}{\partial Y_c} + a_{\zeta} g_{\zeta} - g_{\zeta}^2 \frac{\partial s_d^{\zeta}}{\partial \zeta}.
\end{equation}
This set of equations is closed by the specification of the projections of the flame displacement velocity in $Y_c$ and $\zeta$ directions, which can be obtained from Eq.~(\ref{sd-99}).

A second issue associated with the adoption of the ($Y_c, \zeta$)-space is the fact that the progress variable can present non-monotonic profiles even for simple flames (such as classical non-premixed flames). While this represents an important challenge if the flamelet equations are to be solved in this space directly, the concepts introduced in this work provide an alternative path for their exploration. More specifically, since the ($Z,Y_c$), ($Z,\varphi$) and ($Y_c, \zeta$) spaces are all equivalent (characterized by the same generalized flame displacement velocity), the flamelet equations can be solved in any of these two-dimensional spaces and the results can be then projected into the other two. In this way, the specific closure issues of these coordinate systems can be avoided and variables difficult to model, such as the cross scalar dissipation rates appearing with the adoption of a non-orthogonal coordinate system, become a result of the computations.

Since the entire analysis presented here can be easily reproduced for other sets of non-orthogonal coordinates typically employed in the literature (e.g. two mixture fractions~\citep{Hasse05}, mixture fraction and enthalpy~\citep{Mittal12}, etc.), it is clear that the current approach has opened an entire set of new coordinate systems to be explored in future work.

\section{Conclusions}

In this work, a new flamelet paradigm has been developed, which makes effective use of the Lagrangian nature of the flamelet transformation. The formalism utilizes a generic non-orthogonal coordinate system and it is built around the generalized flame displacement velocity, $\mathbf{s}_d$ (vector), which represents an extension of the classical flame displacement speed, $s_d$ (scalar). It has been shown how this framework allows the derivation of a very compact set of flamelet equations, which facilitates the interpretation of important aspects of flamelet theory. In this sense, a major achievement of the present work has been relating the different terms appearing in the conditioning scalar gradient equations to the generalized flame displacement velocity, which provided important new insights into their physical meaning. Finally, it has been shown how the adoption of $\mathbf{s}_d$ as a central element in flamelet theory allows connecting different non-orthogonal and orthogonal coordinate systems, which opened an entire spectrum of possibilities regarding the definition of new two-dimensional orthogonal composition spaces beyond the different non-orthogonal coordinate systems that have been reported in the literature so far. This new flamelet paradigm complements very well the classical flamelet theory and, at the same time, provides a very suitable basis for future developments.

\appendix 
\numberwithin{equation}{section}
\begin{appendices}%
\renewcommand{\appendixname}{Appendix}
\section{The $\psi$-equation in classical flamelet form}
\label{appendix1}
\renewcommand{\appendixname}{}

It is shown now how Eq.~(\ref{flamelet}) can be brought into a classical flamelet form directly allowing the recovery of the different sets of flamelet equations available in the literature. For this, the equation is first specialized for the generic reactive scalar, $\psi$, which yields
\begin{equation}
     \frac{\partial \psi}{\partial \tau} =  \Gamma_{\psi} -  \Gamma_{\alpha} \frac{\partial \psi}{\partial \alpha} -  \Gamma_{\beta} \frac{\partial \psi}{\partial \beta}.
     \label{flamelet-psi}
\end{equation}
Assuming all $\Gamma$-terms are governed by Eq.~(\ref{gamma}), we can apply the flamelet transformation, Eq.~(\ref{transf}), to $\Gamma_{\psi}$ to obtain
\begin{align}
     \Gamma_{\psi} & =  D g_{\alpha}^2 \frac{\partial^2 \psi}{\partial \alpha^2} +  D g_{\beta}^2 \frac{\partial^2 \psi}{\partial \beta^2} + \frac{\dot{S}_{\psi}}{\rho} \nonumber \\ & + \frac{1}{\rho} \nabla \cdot (\rho D \nabla \alpha) \frac{\partial \psi}{\partial \alpha} + \frac{1}{\rho} \nabla \cdot (\rho D \nabla \beta) \frac{\partial \psi}{\partial \beta} \nonumber \\ & + 2 D g_{\alpha} g_{\beta} \Theta \frac{\partial^2 \psi}{\partial \alpha \partial \beta}.
    \label{rhogammapsi}
\end{align}
Inserting this expression in Eq.~(\ref{flamelet-psi}), multiplying by $\rho$ and taking into account the definition of $\Gamma_{\alpha}$ and $\Gamma_{\beta}$ given by Eq.~(\ref{gamma}),
the following general flamelet equation for $\psi$ can be obtained
\begin{align}
    \rho \frac{\partial \psi}{\partial \tau} & = \rho D g_{\alpha}^2 \frac{\partial^2 \psi}{\partial \alpha^2} + \rho D g_{\beta}^2 \frac{\partial^2 \psi}{\partial \beta^2} + \dot{S}_{\psi} \nonumber \\ & + 2 \rho D g_{\alpha} g_{\beta} \Theta \frac{\partial^2 \psi}{\partial \alpha \partial \beta}  - \dot{S}_{\alpha}\frac{\partial \psi}{\partial \alpha} - \dot{S}_{\beta} \frac{\partial \psi}{\partial \beta}.
    \label{psi-flamelet}
\end{align}
In this form, the connection between the current Lagrangian derivation and the different formulations available in the literature becomes evident. For example, setting $\alpha = Z$ and $\beta = Y_c$, Eq.~(\ref{psi-flamelet}) directly leads to the formulation presented by~\cite{Nguyen10} (see Eq.~(21) in that work). Similarly, adopting $\alpha = Z$ and $\beta = \varphi$,  where $\varphi$ is a modified reaction progress variable satisfying $\nabla Z \cdot \nabla \varphi = 0$ (and therefore implying that $\Theta = 0$),  Eq.~(\ref{psi-flamelet}) reduces to the recent formulation by~\cite{Olguin232} (see Eq.~(41) in that work). Also, it is noted that considering the limit $\partial (\cdot) / \partial \beta \to 0$ and setting either $\alpha = Z$ or $\alpha = Y_c$, the canonical one-dimensional non-premixed equations developed by~\cite{Peters84} and the most recent equivalent premixed equations derived by~\cite{Lodier11} are recovered, respectively. The procedure presented in this section can be easily extended for non-unity Lewis number situations.

\end{appendices}

%aqui referencias
%\bibliographystyle{unsrt}
\bibliography{references}

\begin{thebibliography}{10}
\expandafter\ifx\csname url\endcsname\relax
  \def\url#1{\texttt{#1}}\fi
\expandafter\ifx\csname urlprefix\endcsname\relax\def\urlprefix{URL }\fi
\expandafter\ifx\csname href\endcsname\relax
  \def\href#1#2{#2} \def\path#1{#1}\fi

\bibitem{Peters84}
N.~Peters, Laminar diffusion flamelet models in non-premixed turbulent combustion, Prog. Energy Combust. Sci. 10 (1984) 319 -- 339.

\bibitem{Pitsch98}
H.~Pitsch, N.~Peters, A consistent flamelet formulation for non-premixed combustion considering differential diffusion effects, Combust. Flame 114 (1998) 26 -- 40.

\bibitem{Oijen00}
J.~A. van Oijen, L.~P.~H. de~Goey, Modelling of premixed laminar flames using flamelet-generated manifolds, Combust. Sci. Technol. 161 (2000) 113--137.

\bibitem{Oijen01}
J.~A. van Oijen, F.~A. Lammers, L.~P.~H. de~Goey, Modeling of complex premixed burner systems by using flamelet-generated manifolds, Combust. Flame 127 (2001) 2124 -- 2134.

\bibitem{Lodier11}
G.~Lodier, L.~Vervisch, V.~Moureau, P.~Domingo, Composition-space premixed flamelet solution with differential diffusion for in situ flamelet-generated manifolds, Combust. Flame 158 (2011) 2009 -- 2016.

\bibitem{vanOijen16}
J.~A. van Oijen, A.~Donini, R.~J.~M. Bastiaans, J.~H.~M. ten Thije~Boonkkamp, L.~P.~H. de~Goey, State-of-the-art in premixed combustion modeling using flamelet generated manifolds, Prog. Energy Combust. Sci. 57 (2016) 30--74.

\bibitem{Xu13}
H.~Xu, F.~Hunger, M.~Vascellari, C.~Hasse, A consistent flamelet formulation for a reacting char particle considering curvature effects, Combust. Flame 160 (2013) 2540 -- 2558.

\bibitem{Xuan14}
Y.~Xuan, G.~Blanquart, M.~E. Mueller, Modeling curvature effects in diffusion flames using a laminar flamelet model, Combust. Flame 161 (2014) 1294 -- 1309.

\bibitem{Olguin13}
H.~Olguin, E.~Gutheil, Influence of evaporation on spray flamelet structures, Combust. Flame 161 (2014) 987 -- 996.

\bibitem{Olguin14}
H.~Olguin, E.~Gutheil, Theoretical and numerical study of evaporation effects in spray flamelet modeling, in: B.~Merci, E.~Gutheil (Eds.), Experiments and Numerical Simulations of Turbulent Combustion of Diluted Sprays, Vol.~19 of ERCOFTAC Series, Springer International Publishing, 2014, pp. 79--106.

\bibitem{Scholtissek15}
A.~Scholtissek, W.~L. Chan, H.~Xu, F.~Hunger, H.~Kolla, J.~H. Chen, M.~Ihme, C.~Hasse, A multi-scale asymptotic scaling and regime analysis of flamelet equations including tangential diffusion effects for laminar and turbulent flames, Combust. Flame 162 (2015) 1507 -- 1529.

\bibitem{Olguin15}
H.~Olguin, E.~Gutheil, Derivation and evaluation of a multi-regime spray flamelet model, Zeitschrift für Physikalische Chemie 229~(4) (2015) 461--482.

\bibitem{Scholtissek17}
A.~Scholtissek, F.~Dietzsch, M.~Gauding, C.~Hasse, In-situ tracking of mixture fraction gradient trajectories and unsteady flamelet analysis in turbulent non-premixed combustion, Combust. Flame 175 (2017) 243 -- 258.

\bibitem{Scholtissek18}
A.~Scholtissek, P.~Domingo, L.~Vervisch, C.~Hasse, A self-contained progress variable space solution method for thermochemical variables and flame speed in freely-propagating premixed flamelets, Proc.~Combust.~Inst. 37 (2018) 1529 -- 1536.

\bibitem{Scholtissek19}
A.~Scholtissek, P.~Domingo, L.~Vervisch, C.~Hasse, A self-contained composition space solution method for strained and curved premixed flamelets, Combust. Flame 207 (2019) 342--355.

\bibitem{Olguin19}
H.~Olguin, A.~Scholtissek, S.~Gonzalez, F.~Gonzalez, M.~Ihme, C.~Hasse, E.~Gutheil, Closure of the scalar dissipation rate in the spray flamelet equations through a transport equation for the gradient of the mixture fraction, Combust. Flame 208 (2019) 330--350.

\bibitem{Olguin23}
H.~Olguin, F.~Huenchuguala, Z.~Sun, C.~Hasse, A.~Scholtissek, Three questions regarding scalar gradient equations in flamelet theory, Combust. Flame 249 (2023) 112624.

\bibitem{Hasse05}
C.~Hasse, N.~Peters, A two mixture fraction flamelet model applied to split injections in a {DI} diesel engine, Proc. Combust. Inst. 30 (2005) 2755 -- 2762.

\bibitem{Domingo08}
P.~Domingo, L.~Vervisch, D.~Veynante, Large-{E}ddy {S}imulation of a lifted methane jet flame in a vitiated coflow, Combust. Flame 152 (2008) 415 -- 432.

\bibitem{Knudsen09}
E.~Knudsen, H.~Pitsch, A general flamelet transformation useful for distinguishing between premixed and non-premixed modes of combustion, Combust. Flame 156 (2009) 678 -- 696.

\bibitem{Nguyen10}
P.-D. Nguyen, L.~Vervisch, V.~Subramanian, P.~Domingo, Multidimensional flamelet-generated manifolds for partially premixed combustion, Combust. Flame 157 (2010) 43 -- 61.

\bibitem{Knudsen12}
E.~Knudsen, H.~Pitsch, Capabilities and limitations of multi-regime flamelet combustion models, Combust. Flame 159 (2012) 242 -- 264.

\bibitem{Mittal12}
V.~Mittal, D.~J. Cook, H.~Pitsch, An extended multi-regime flamelet model for {IC} engines, Combust. Flame 159 (2012) 2767 -- 2776.

\bibitem{Mueller20}
M.~E. Mueller, Physically-derived reduced-order manifold-based modeling for multi-modal turbulent combustion, Combust. Flame 214 (2020) 287 -- 305.

\bibitem{Scholtissek20}
A.~Scholtissek, S.~Popp, S.~Hartl, H.~Olguin, P.~Domingo, L.~Vervisch, C.~Hasse, Derivation and analysis of two-dimensional composition space equations for multi-regime combustion using orthogonal coordinates, Combust. Flame 218 (2020) 205 -- 217.

\bibitem{Novoselov21}
A.~G. Novoselov, B.~A. Perry, M.~E. Mueller, Two-dimensional manifold equations for multi-modal turbulent combustion: Nonpremixed combustion limit and scalar dissipation rates, Combust. Flame 231 (2021) 111475.

\bibitem{Olguin232}
H.~Olguin, P.~Domingo, L.~Vervisch, C.~Hasse, A.~Scholtissek, A self-consistent extension of flamelet theory for partially premixed combustion, Combust. Flame 255 (2023) 112911.

\bibitem{Olguin24}
H.~Olguin, P.~Domingo, L.~Vervisch, C.~Hasse, A.~Scholtissek, On the closure of curvature in 2{D} flamelet theory, Combust. Flame 267 (2024) 113599.

\bibitem{Gibson68}
C.~H. Gibson, Fine structure of scalar fields mixed by turbulence. i. zero gradient points and minimal gradient surfaces, Phys. Fluids 11 (1968) 2305--2315.

\bibitem{Williams85}
F.~Williams, Combustion Theory, Oxford University Press, 1985.

\bibitem{Poinsot05}
T.~Poinsot, D.~Veynante, Theoretical and Numerical Combustion, Second Edition, R.T. Edwards, Inc., 2005.

\bibitem{Echekki99}
T.~Echekki, J.~H. Chen, Analysis of the contribution of curvature to premixed flame propagation, Combust. Flame 118~(1) (1999) 308--311.

\bibitem{Peters00}
N.~Peters, Turbulent Combustion, Cambridge University Press, 2000.

\bibitem{Pitsch982}
H.~Pitsch, M.~Chen, N.~Peters, Unsteady flamelet modeling of turbulent hydrogen-air diffusion flames, Symposium (International) on Combustion 27 (1998) 1057 -- 1064, twenty-Seventh Sysposium (International) on Combustion Volume One.

\bibitem{Pitsch00}
H.~Pitsch, Unsteady flamelet modeling of differential diffusion in turbulent jet diffusion flames, Combust. Flame 123 (2000) 358 -- 374.

\bibitem{Klimenko01}
A.~Y. Klimenko, On the relation between the conditional moment closure and unsteady flamelets, Combust. Theory Model. 5 (2001) 275--294.

\bibitem{Mantel94}
T.~Mantel, R.~Borghi, A new model of premixed wrinkled flame propagation based on a scalar dissipation equation, Combust. Flame 96 (1994) 443 -- 457.

\bibitem{Kollmann98}
W.~Kollmann, J.~H. Chen, Pocket formation and the flame surface density equation, Symp. (Int.) Combust. 27~(1) (1998) 927--934, twenty-Seventh Sysposium (International) on Combustion Volume One.

\bibitem{Chakraborty05}
N.~Chakraborty, R.~Cant, Effects of strain rate and curvature on surface density function transport in turbulent premixed flames in the thin reaction zones regime, Phys. Fluids 17~(6) (2005) 065108.

\bibitem{Swaminathan05}
N.~Swaminathan, K.~Bray, Effect of dilatation on scalar dissipation in turbulent premixed flames, Combust. Flame 143~(4) (2005) 549--565, special Issue to Honor Professor Robert W. Bilger on the Occasion of His Seventieth Birthday.

\bibitem{Chakraborty08}
N.~Chakraborty, M.~Klein, Influence of lewis number on the surface density function transport in the thin reaction zone regime for turbulent premixed flames, Phys. Fluids 20~(6) (2008) 065102.

\bibitem{Dopazo16}
C.~Dopazo, L.~Cifuentes, The physics of scalar gradients in turbulent premixed combustion and its relevance to modeling, Combust. Sci. Technol. 188~(9) (2016) 1376--1397.

\bibitem{Sandeep18}
A.~Sandeep, F.~Proch, A.~M. Kempf, N.~Chakraborty, {Statistics of strain rates and surface density function in a flame-resolved high-fidelity simulation of a turbulent premixed bluff body burner}, Phys. Fluids 30~(6) (2018) 065101.

\bibitem{Cifuentes18}
L.~Cifuentes, C.~Dopazo, A.~Sandeep, N.~Chakraborty, A.~Kempf, {Analysis of flame curvature evolution in a turbulent premixed bluff body burner}, Phys. Fluids 30~(9) (2018) 095101.

\bibitem{Dopazo18}
C.~Dopazo, L.~Cifuentes, D.~Alwazzan, N.~Chakraborty, Influence of the lewis number on effective strain rates in weakly turbulent premixed combustion, Combust. Sci. Technol. 190~(4) (2018) 591--614.

\bibitem{Yu19}
R.~Yu, T.~Nillson, X.-S. Bai, A.~N. Lipatnikov, Evolution of averaged local premixed flame thickness in a turbulent flow, Combust. Flame 207 (2019) 232--249.

\bibitem{Yu192}
R.~Yu, A.~N. Lipatnikov, Surface-averaged quantities in turbulent reacting flows and relevant evolution equations, Phys. Rev. E 100 (2019) 013107.

\bibitem{Peters09}
N.~Peters, Multiscale combustion and turbulence, Proc. Combust. Inst. 32 (2009) 1 -- 25.

\bibitem{DeGoey97}
L.~P.~H. de~Goey, J.~H.~M. ten Thije~Boonkkamp, A mass-based definition of flame stretch for flames with finite thickness, Combust. Sci. Technol. 122 (1997) 399--405.

\bibitem{DeGoey99}
L.~P.~H. de~Goey, J.~H.~M. ten Thije~Boonkkamp, A flamelet description of premixed laminar flames and the relation with flame stretch, Combust. Flame 119 (1999) 253--271.

\end{thebibliography}

\end{document}